Report on the ESO Workshop

# The Present and Future of Astronomy (ASTRO2022)

held online, 14–18 February 2022

Giacomo Beccari[1]
Henri M. J. Boffin[1]
Paola Andreani[1]
Selma de Mink[2]
Wendy Freedman[3]
Michael Hill[4]
Bruno Leibundgut[1]
Federico Lelli[5]
Anna Miotello[1]
Sean Sapcariu[6]

[1] ESO
[2] Max Planck Institute for Astrophysics, Garching, Germany
[3] University of Chicago, USA
[4] Swiss National Science Foundation, Switzerland
[5] INAF Florence, Italy
[6] Fonds National de la Recherche, Luxembourg

| | | |
|---|---|---|
| Monday, 14 February 2022 | Methodology of science in the modern world | |
| Tuesday, 15 February 2022 | The funding of astronomy | |
| Wednesday, 16 February 2022 | Assessment and metrics | |
| Thursday, 17 February 2022 | Mental health and impostor syndrome | |
| Friday, 18 February 2022 | Astronomy and society | |

Table 1. The programme of the ASTRO2022 conference.

Being one of the most fascinating and ancient sciences, astronomy has always played a special role in society. In 2022 ESO organised an online conference to offer the community a platform to discuss astronomical topics of sociological and philosophical relevance in a professional atmosphere. The talks touched on several crucial aspects, moving from the methodology of science to the use of metrics, to the importance of diversity in evaluation processes, and to the link between astronomy and society.

Science plays a crucial role in modern society. Scientists foster knowledge and produce results that have often immediate impact on people's lives. In this context, astronomy plays a privileged role. While having an extraordinary impact on the imagination of many people, astronomy enables researchers to use cutting-edge technology to explore the Universe and its beauty. In this way, astronomers can play a crucial role in science education. The development of large, international and collaborative organisations like ESO and their research facilities at the forefront of technology brings many broader societal benefits[1].

Driven by a genuine passion for their work, the community of professional astronomers is permeated by a continuous exchange of technical expertise, basic knowledge and scientific discoveries that crosses most cultural barriers. Diversity, personal development and returns to society are important aspects of our profession too.

Astronomical research is communicated mostly through well-established networks of professional journals, conferences, workshops, public presentations and teaching. Scientific publications play a crucial role not only in the development and sharing of knowledge but also in the sustainability of the research activity itself. It is not uncommon practice to use publication-related metrics in evaluation, assessments of performance and hiring processes. This fact has in recent years put significant pressure on researchers at all career stages, with the risk that the productivity of each individual might gain more relevance than the quality of the research.

ESO organised an online workshop where the astronomical community was invited to discuss, with the help of professionals from other disciplines, including social sciences, the above points and potential for improvements. Driven by the positive experience of the ESO Cosmic Duologues and Hypatia series (Beccari & Boffin, 2020, 2021), we have now learned to what extent online seminars can be an efficient way to engage and reach out to the astronomical community, even under the severe restrictions imposed by the COVID-19 pandemic.

The workshop (ASTRO2022) was announced in November 2021. The community was invited to participate either by registering for the conference or by following the event live on YouTube. By the deadline on the 9 February 2022, 486 participants had registered. Of the registered participants, 52% were early-career scientists (35% PhD students and 17% postdoc). This clearly demonstrates the strong interest within the astronomical community as a whole, and particularly the youngest, in participating in a workshop critically assessing how science is done in the modern era. It is also important to note that 45% of the registered participants were female. While it is not easy to demonstrate whether these numbers are also representative of the effective demographic of the participants in the online event, we are proud to acknowledge that we achieved our goal of engaging a diverse group of people.

## Five conferences in one: the programme

As said by several participants, this was a unique workshop that addressed essential themes of concern for every researcher. Each day was dedicated to a particular topic: Methodology of science in the modern world; The funding of astronomy; Assessment and metrics; Mental health and impostor syndrome; Astronomy and society. It was said several times that each of these topics would have deserved a dedicated conference of its own. The videos of the event are available on YouTube[2] and some of the presentations have been made publicly available on Zenodo[3]. At the time of writing (March 2022), the videos have been viewed more than 3700 times.

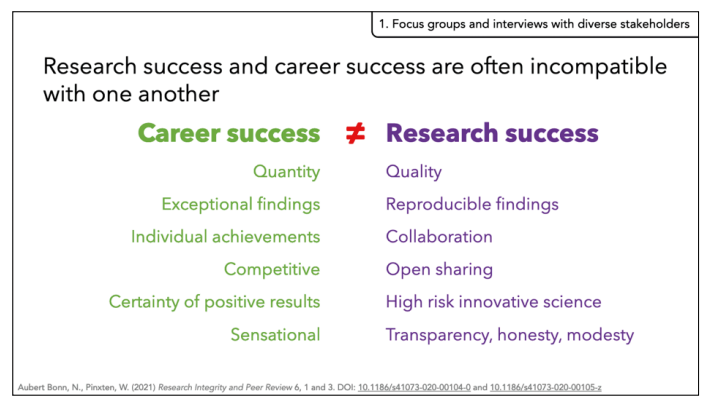

Figure 1. A slide from the presentation of Noémie Aubert Bonn showing the difference between the words characterising career and research success in the current times.

In his introduction, Bruno Leibundgut from ESO underlined some of the key aspects that motivated the meeting: as professional astronomers, we are privileged to do the work we do and honoured by the trust that society places in us by investing research funds to create knowledge. It is a responsibility of the research community to reflect and take a critical look at the various aspects that crucially impact our work, like the use of metrics in hiring and in assessing, the importance of ensuring diversity to guarantee a healthy scientific environment, the use of funding, and the role of astronomy in education.

The conference started with a series of talks touching on the foundations of the scientific method from a philosophical perspective. It was remarkable to see almost 300 participants connected via either YouTube or Zoom during the first day. Given the large participation, there is no doubt that our community is willing to understand whether we still follow the fundamental principles of the scientific method as stated by Karl Popper in the last century. And the answer is certainly far from trivial. It is important, however, to realise that such studies exist, and that as researchers we need to take a step back and analyse our methods.

On the second day, the topic of the conference moved to the funding of astronomical research. Valentin Oprea described the opportunities provided by the European Research Council, while Thomas Zurbuchen highlighted the importance of ensuring diversity in hiring and funding at NASA. This last point was also the focus of the contributed talks and the discussion, reinforcing the need for greater diversity — in gender, professional profile, culture, and minorities, amongst others — in shaping healthy and successful scientific programmes. As explained by Francesca Primas, member of the ESO Diversity and Inclusion Committee, diversity is one of the ESO organisational goals and is present in various critical aspects, from playing a key role in shaping the Respectful Workplace policies aimed at fostering professional and respectful relationships at work, to the formulation of mid- and long-term goals to consolidate diversity and inclusion in recruitment and promotion processes.

The discussion on recruitment and assessment processes continued on the third day of the conference when, thanks to the talks by Stephen Curry and Noémie Aubert Bonn, the focus was the use of metrics. Noémie Aubert Bonn presented the results of a study aimed at identifying the perceived meaning of scientific and career success in a research environment. As shown quite remarkably in Figure 1, the keywords describing research success and career success are often incompatible. While the availability of large and complete databases has made it possible to perform accurate metric analyses in evaluation processes, it is a fact that the first modern concept of peer review was introduced at least two centuries ago, when the number of humans populating the planet was a factor of sixteen lower than it is today (see Figure 2). It is therefore a fair question to ask whether the peer review system as originally introduced still allows a solid and balanced evaluation of the scientific quality of manuscripts, and whether the introduction of metrics alleviates or exacerbates biases in the evaluation process. The San Francisco Declaration on Research Assessment (the DORA initiative[4]) presented by Stephen Curry, looks critically at these and other aspects of the evaluation process and aims to find new means to renovate the ways in which the outputs of scholarly research are evaluated and adapt to the requirements of a modern world (for example, by eliminating the use of journal-based metrics, and assessing research on its own merits).

The restrictions imposed by the COVID-19 pandemic had, in the course of 2020 and 2021, dramatic consequences for career development, especially for early-career scientists. The delicate balance between professional and personal life has been challenged by imposed social distancing, lockdown measures and isolation. Mental health and wellbeing have become more than ever crucial aspects of our personal and professional life which deserve awareness within the community. Ewa Pluciennicka and Sanne Feenestra presented the latest scientific insights into mental health and the impostor phenomenon in academia. As emphasised by the valuable contributions from all the Thursday speakers, it is imperative that institutions listen to the voices of the new generation of astronomers whose desires, needs and vision often challenge the established procedures and approaches in research. Family support, work-life balance, employment of personnel with disabilities, and support of researchers in developing countries are only a few of the aspects that must become the core of scientific and research policies[5]. As long as the growth of scientific knowledge and the horizon of a career in research are dominated by productivity and associated concepts like the number of publications and leadership, we risk forgetting what lies at the core of scientific progress: the individual scientist, their creativity and talent.

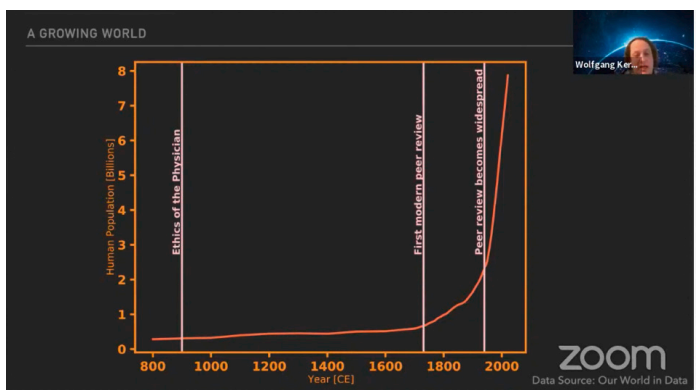

Figure 2. A slide from Wolfgang Kerzendorf's presentation showing the increase of human population as a function of time. The vertical lines indicate three milestones on the road to the modern peer review process.

There are extraordinary and successful examples of how astronomy can be used to reach the general public, inspire the youngest generation and contribute to bringing more equality into the world. Sandra Benítez Herrera and Mirjana Povič showed how initiatives related to research collaborations, education, institutional development, human capacity building, policy development, and participation of women in science are powerful ways to stimulate creativity and increase the perception of science as an approachable discipline contribute to combat poverty and inequity in the world.

### Closing remarks: more than "food for thought"

The ASTRO2022 conference offered to many participants a starting point to address, in a fast-changing world, important questions about the foundations of the scientific method and the modernity of science. Scientists have to ask themselves what their responsibility is and — in our context — what we, as professional astronomers, can do to exert a positive influence on society. The topics discussed in the five half-days touched on aspects that we all need to understand and accept. Diversity in science, wellbeing and work-life balance, sustainability, fair assessments, and quality-driven evaluations are aspects to be fully ingested and represented in science- and research-related policies and procedures. There is a profound desire in the scientific community and beyond to refocus our efforts onto people. A fair society in which every person can flourish based on their own talents, desires and creativity must be the goal of the changes in the way we do science in a modern world. The talks and discussions triggered during the ASTRO2022 conference were a promising start towards a new view of science in a modern world. Achieving this ambitious goal requires fair and open inter-generational exchanges with the support of multi-disciplinary expertise.

### Acknowledgements

We would like to thank all the speakers and the community for their enthusiastic and professional participation in the live event and in the discussion sessions.

### References

Beccari, G. & Boffin, H. M. J. 2020, The Messenger, 181, 34
Beccari, G. & Boffin, H. M. J. 2021, The Messenger, 185, 23

### Links

[1] ESO's Benefits to Society publication: https://www.eso.org/public/products/brochures/brochure_0076/
[2] ASTRO2022 videos: https://www.youtube.com/c/ESOCosmicDuologues
[3] ASTRO2022 presentations: https://zenodo.org/communities/astro2022
[4] DORA: https://sfdora.org/
[5] For examples on work-family balance and issues affecting especially mothers, see the book "Mothers in Astronomy" by Paola Pinilla, https://misaladino.com/mothers-in-astronomy/

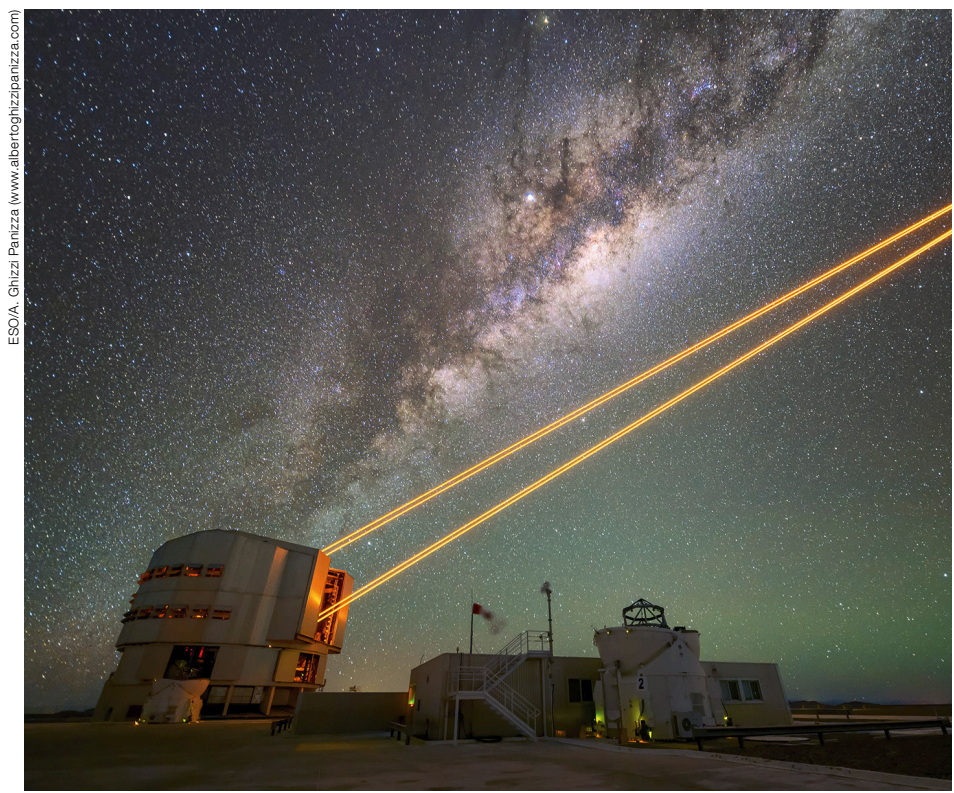

This picture of the week shows four powerful laser beams leaving Unit Telescope 4, or 'Yepun', at ESO's Very Large Telescope (VLT) in Chile's Atacama desert. These form the VLT's 4 Laser Guide Star Facility, which enables astronomers to take extremely crisp images of the cosmos by employing a technology known as adaptive optics.